\documentclass[authoryear,12pt,a4paper,times]{elsarticle}

\usepackage{algorithmic}
\usepackage{algorithm}

\usepackage{geometry}
\usepackage{amssymb}

\usepackage{setspace}

\usepackage[figuresright]{rotating}

\begin{document}

\begin{frontmatter}



\title{Improving Lower Bounds for the Quadratic Assignment Problem by  applying a  Distributed Dual Ascent Algorithm}

\author[adg]{A. D. Gon\c calves}
\ead{agoncalves@ic.uff.br}

\author[adg]{L. M. A. Drummond \corref{cor1}}
\ead{lucia@ic.uff.br}

\author[aap]{A. A. Pessoa}
\ead{artur@producao.uff.br}

\author[pmh]{P. M. Hahn}
\ead{hahn@seas.upenn.edu}

\cortext[cor1]{Corresponding author: Tel.: +55  21  26295680; Fax: +55  21  26295667}


\address[adg]{Computer Science Department, Fluminense Federal University, Niter\'{o}i, RJ, Brazil}

\address[aap]{Production Engineering Department, Fluminense Federal University, Niter\'{o}i - RJ, Brazil}

\address[pmh]{Electrical and Systems Engineering, The University of Pennsylvania, Philadelphia, PA 19104-6315, USA}

\begin{abstract}
The application of the Reformulation Linearization Technique (RLT) to the Quadratic Assignment Problem (QAP) leads to a tight linear relaxation
with huge dimensions that is hard to solve.
Previous works found in the literature show that these relaxations combined with branch-and-bound algorithms belong to the state-of-the-art
of exact methods for the QAP. 
For the level 3 RLT (RLT3), using this relaxation is prohibitive in conventional machines for instances with more than 22 locations
due to memory limitations.
This paper presents a distributed version of a dual ascent algorithm for the RLT3 QAP relaxation that approximately solves it for instances
with up to 30 locations for the first time. Although, basically, the  distributed algorithm
has been  implemented on top of its  sequential conterpart, some changes,
 which improved not only the parallel performance
but  also the quality of solutions, were proposed here.
When compared to other lower bounding methods found in the literature, our algorithm generates the best known lower bounds for 26 out of the
28 tested instances, reaching the optimal solution in 18 of them.
\end{abstract}


\end{frontmatter}

\onehalfspacing 

\section{Introduction} \label{sec:introducao}

Given $N$ objects, $N$ locations, a flow $f_{ik}$ from each object $i$ to each object $k$, $k \neq i$, and a distance $d_{jn}$ from each location $j$ to each location
$n$, $n \neq j$, the quadratic assignment problem (QAP) consists of assigning  each object $i$ to exactly  a location $j$. We wish to find:

\begin{equation}
\label{eq:n1} min  \displaystyle  \sum_{i=1}^{N} \sum_{j=1}^{N} \sum_{k=1 \atop {k\not=i} }^{N} \sum_{n=1 \atop {n\not=j}}^{N} f_{ik} d_{jn} x_{ij} x_{kn} : x \in X , \  x \in  \left\{0,1\right\} \\
\end{equation}


Initially presented by \citet{koopman1957}, 
the QAP has practical applications in  several areas, such as facility layout, electronic circuit board design, construction planning, etc.  
The QAP is one of the most difficult and studied combinatorial optimization problems found in OR literature. Usually, difficult instances require 
a great deal of computational effort to be solved exactly. In \citet{adams2007},  for example, a 30-location instance is solved on a single 
cpu of a Dell 7150 PowerEdge server in 1,848 days. 
Thus, good lower bounds are crucial for solving  instances with more than 15 
locations in reasonable processing time.  They would allow that  a large number of  alternative solutions  is discarded  during  the search for the  optimal solution in the branch-and-bound tree. 

A summary of the techniques used for calculating lower bounds  is presented in \citet{loiola2007}. 
In the QAPLIB website \citet{qaplib}, a table showing lower bounds for each instance of the site is presented.
The best lower bounds were achieved by \citet{burer2006}, \citet{adams2007} and  \citet{hahn2012}.
The dual ascent algorithm based on the RLT3 formulation, described in \citet{hahn2012}, 
calculates tight lower bounds, but the use of such technique in conventional machines for instances with more than 25 locations is impossible due to 
its large memory requirements. For example, to solve an instance of 25 locations, Hahn, in \citet{hahn2012}, used a host with 173 GB of shared memory.
Recently, a very difficult instance with 30 locations has been solved exactly also using the RLT3 formulation (see
http://www.seas.upenn.edu/qaplib/news.html). In this case, the authors used the same cluster of machines, which contains hosts with up to 2 TB
of shared memory. 

The contribution of this paper is the proposal of a distributed  application developed
on top of  the   sequential  algorithm proposed in \citet{hahn2012},  but not equivalent to it, since our new algorithm has some important changes, which improve not only the performance
but  also the quality of RLT3 lower bounds for some instances.   This distributed algorithm  executes  on  a conventional  cluster of computers  and generates the best known lower bounds for 26 out of the
28 tested instances, reaching the optimal solution in 18 of them.

\section{Reformulation-linearization technique applied to the QAP} \label{sec:rlt}

The reformulation-linearization technique was initially developed by  
\citet{adams1986}, aiming to generate tight linear programming relaxations for discrete and continuous nonconvex problems.
For mixed zero-one programs involving $m$ binary variables, RLT establishes an $m$-level hierarchy of relaxations spanning from the ordinary linear programming relaxation 
to the convex hull of feasible integer solutions. For a given   $z \in \{i,..,m\}$, the level-$z$ RLT, or RLT$z$, constructs various polynomial factors of degree $z$
consisting of the product of some $z$ binary variables $x_{j}$ or their complements $(1 - x_{j})$. We find in the literature various RLT levels applied 
to the QAP, RLT1 in \citet{hahn1998}, RLT2 in \citet{adams2007} and RLT3 in \citet{hahn2012}. The RLT consists of two steps: the reformulation and linearization. 

The RLT3 reformulation, presented in \citet{hahn2012}, consists of the following steps: (i) multiply each of $2N$ assignment constraints by each of the $N^2$ 
binary variables $x_{ij}$ (applying  RLT1); (ii) multiply each of the $2N$ assignment constraints by each one of the $N^2(N-1)^2$ 
products $x_{ij}x_{kn}$, $k\not=i\ $ and $\ n\not=j$ (applying RLT2); (iii) multiply each of the $2N$ assignment constraints by each one of the
 $N^2(N-1)^2(N-2)^2$ products $x_{ij}x_{kn}x_{pq}$,  $p\not=k\not=i\ $ and $q\not=n\not=j$ (applying RLT3). Moreover, remove the products $x_{ij}x_{kn}$ 
if  $(k=i\ $and$\ n\not=j)\ $or$\ (k\not=i\ $and$\ n=j)$ in quadratic expressions;
remove all products  $x_{ij}x_{kn}x_{pq}$ if $(p=i\ $and$\ q\not=j),\ (p=k\ $ and $\ q\not=n),\ (p\not=i\ $and$\ q=j) $ or $ (p\not=k\ $and$\ q=n)$ in cubic expressions;  
and, finally,  remove all products $x_{ij}x_{kn}x_{pq}x_{gh}$ if $(g=i\ $and$\ h\not=j),\ (g=k\ $and$\ h\not=n),\ (g=p\ $and$\ h\not=q),\ (g\not=i\ $and$\ h=j),\ 
(g\not=k\ $and$\ h=n)\ $ or $\ (g\not=p\ $and$\ h=q)$ in biquadratic expressions. 

The linearization consists of: (i) replace each product $x_{ij}x_{kn}$, with $i\not=k$ and $j\not=n$, by the continuous variable $y_{ijkn}$,   
imposing the constraints  $y_{ijkn} = y_{knij}$ (2 complementaries) for all $(i,j,k,n)$ with $i<k$ and $j\not=n$ (applying RLT1); (ii) 
replace each product $x_{ij}x_{kn}x_{pq}$, with $i\not=k\not=p$ and $j\not=n\not=q$, by the continuous variable $z_{ijknpq}$, imposing the constraints 
$z_{ijknpq}=z_{ijpqkn}=z_{knijpq}=z_{knpqij}=z_{pqijkn}=z_{pqknij}$ (6 complementaries) for all $(i,j,k,n,p,q)$ with $i<k<p$ and $j\not=n\not=q$  (applying RLT2);
(iii) replace each product $x_{ij}x_{kn}x_{pq}x_{gh}$ for $v_{ijknpqgh}$, with $i\not=k\not=p\not=g$ and $j\not=n\not=q\not=h$, by the continuous variable $v_{ijknpqgh}$,
imposing the constraints $v_{ijknpqgh}=v_{ijknghpq}= ...=v_{ghpqknij}$ (24 complementaries) for all $(i,j,k,n,p,q,g,h)$ with $i<k<p<g$ and $j\not=n\not=q\not=h$ (applying RLT3).

At the end of RLT3 reformulation, we achieve the following objective function:

\begin{equation}
\label{eq:n2a}  min \left\{\displaystyle \sum_{i=1}^{N} \sum_{j=1}^{N} B_{ij}x_{ij} +  \sum_{i=1}^{N} \sum_{j=1}^{N} \sum_{k=1 \atop{ k\not= i}}^{N} \sum_{n=1 \atop{n\not= j}}^{N} C_{ijkn} y_{ijkn} + 
   \displaystyle  \sum_{i=1}^{N} \sum_{j=1}^{N} \sum_{k=1 \atop { k\not= i}}^{N} \sum_{n=1 \atop{n\not= j}}^{N} \sum_{p=1 \atop{ p\not= i,k}}^{N} \sum_{q=1 \atop{q\not= j,n}}^{N}D_{ijknpq} z_{ijknpq}
   \atop { \displaystyle + \sum_{i=1}^{N} \sum_{j=1}^{N} \sum_{k=1 \atop{ k\not= i}}^{N} \sum_{n=1 \atop{n\not= j}}^{N} \sum_{p=1 \atop{p\not= i,k}}^{N} \sum_{q=1 \atop{ q\not= j,n}}^{N}  \sum_{g=1 \atop{ g\not= i,k,p}}^{N} \sum_{h=1 \atop{h\not= j,n,q}}^{N}E_{ijknpqgh} v_{ijknpqgh}  + LB    }   \right\}
\end{equation}

\

In the objective function (\ref{eq:n2a}), consider
 the constant term $LB = 0$,
 each coefficient $B_{ij} = 0\ \forall\ (i,j)$, 
 each coefficient $C_{ijkn} = f_{ik} \times d_{jn}\ \forall \ (i,j,k,n)\ $ with $ \ i\not=k$ and $j\not=n$,
 each coefficient $D_{ijknpq} = 0\ \forall\ (i,j,k,n,p,q)\ $ with $  \ i\not=k\not=p$ and $j\not=n\not=q\ $, 
 each coefficient $E_{ijknpqgh} = 0\ \forall\ (i,j,k,n,p,q,g,h) \ $ with $ \ i\not=k\not=p\not=g$ and $j\not=n\not=q\not=h$.

The dual ascent algorithm  proposed in \citet{hahn2012} consists of updating the constant term $LB$
and the cost matrices $B$, $C$, $D$ and $E$
in such a way that the cost of any (integer) feasible solution with respect to the modified objective function
remains unchanged, while maintaining nonnegative coefficients. 
As a consequence of this property, the value of $LB$ at any moment of the execution is a valid
lower bound on the optimal solution cost for the QAP.
In the light of these aspects, the following procedures  are  developed:

\begin{enumerate}
\item [I.]
     \textbf{Cost spreading:} consists of the cost distributions from matrix $B$ to $C$, from matrix $C$ to $D$ and from matrix $D$ to $E$. 
      In the cost spreading procedure  from matrix $B$ to $C$, for each $(i,j)$, the  coefficient $B_{ij}$ is spread
      through $(N - 1)$ rows of matrix  $C$, i.e., 
      each  element $C_{ijkn}$ is added by  $B_{ij}\ /\ (N-1)$, $\forall \ k\not=i\ $ and $\ n\not=j$.  
      After such  updating, $B_{ij}$ is updated to $0$  for each $(i,j)$.
      The same procedure is repeated from matrix $C$ to $D$, where each  coefficient  $C_{ijkn}$ is spread
      through $(N - 2)$  rows of matrix
      $D$, and from matrix $D$ to $E$, where each coefficient $D_{ijknpq}$ is spread through $(N - 3)$ rows of matrix $E$. 
 
\item[II.] 
     \textbf{Cost concentration:}  in this procedure we used the Hungarian Algorithm,  \cite{munkres1957}, to
     concentrate the costs from matrix $E$ to $D$, from matrix $D$ to $C$, from matrix $C$ to $B$ and from matrix $B$ to $LB$.
     The cost concentrations from matrix $E$ to $D$ are represented as $D_{ijknpq}$ $\leftarrow  Hungarian(E_{ijknpq})$.
     This procedure uses a  matrix $M$ with size $(N-3)^2$  to receive the $(N-3)^2$ coefficients of the
     submatrix $E_{ijknpq}$: for each $(r,s = 1,..,N-3)$, $M_{rs}$ receives
     $E_{ijknpqgh}$, where $g$ ($h$) is the $r$-th row ($s$-th column) different from $i,k,p$ ($j,n,q$) in the
     submatrix $E_{ijknpq}$. Then, the Hungarian algorithm is applied to $M$
     to obtain the total cost to be added to $D_{ijknpq}$, and the coefficients of the submatrix $E_{ijknpq}$ are
     replaced by the corresponding residual coefficients from $M$.
	The same procedure is repeated as $C_{ijkn}$ $\leftarrow  Hungarian(D_{ijkn})$,  
	$B_{ij}$ $\leftarrow  Hungarian(C_{ij})$ and $LB \leftarrow  Hungarian(B)$.
	In these procedures,
	the sizes of $M$ are $(N-2)^2$, $(N-1)^2$ and $N^2$, respectively.

\item [III.] 
    \textbf{Costs transfer between complementary coefficients:} 
      Differently from \cite{hahn2012}, the cost transfers always replace each coefficient by
      the arithmetic mean of all its complementaries. It is applied as follows: 
      (i) In the matrix $C$, for each $(i,j,k,n)$,  $C_{ijkn} \leftarrow $  $C_{knij} \leftarrow $  $(C_{ijkn} + C_{knij}) / 2 $,  with $ i<k\ $ and $ j\not=n$;
      (ii) In the matrix $D$, for each $(i,j,k,n,p,q)$, $D_{ijknpq} \leftarrow $  $D_{ijpqkn} \leftarrow $  $D_{knijpq} \leftarrow $  $D_{knpqij} \leftarrow $ 
         $D_{pqijkn} \leftarrow $  $D_{pqknij} = (D_{ijknpq} + $ $D_{ijpqkn} + $  $D_{knijpq} + $  $D_{knpqij} + $  $D_{pqijkn} + $  $D_{pqknij})/6 $, 
         with $ i<k<p\ $ and $ \ j\not=n\not=q$;
      (iii) In the matrix $E$, for each $(i,j,k,n,p,q,g,h)$, $E_{ijknpqgh}$ $\leftarrow$ $E_{ijknghpq}$ $\leftarrow$ ... $\leftarrow$ 
         $E_{ghpqknij}$ $\leftarrow$ $(E_{ijknpqgh} + E_{ijknghpq} +\ ...\ + E_{ghpqknij})/24$, with $ i<k<p<g\ $ and $\ j\not=n\not=q\not=h$.

\end{enumerate}

\section{Distributed Algorithm} \label{sec:modelo}

In our distributed version, consider $T$ the set of hosts running the application, and let $R_t\ (R_t\in T)$ be the identification of a host.
 Let $f_{ik}$ and $d_{jn}$  be  flow and distance matrices respectively, according to equation (\ref{eq:n1}), 
$LB$, the lower bound, and $B$,$C$,$D$ and $E$, the matrices presented in the objective function (\ref{eq:n2a}).
Consider $G_{ij}$ as a set composed of submatrices $B$, $C$, $D$, $E$  with the same $(i,j)$ stored and processed on  $R_t$. 
Sets of  $G$ are evenly distributed among the hosts. See Figure 1 for an example with twenty hosts, running an instance of $N=20$. 
In this figure,  the set $G_{15,7}$ composed of submatrices $B_{15,7}$,  $C_{15,7,k,n}$,  $D_{15,7,k,n,p,q}$ and 
$E_{15,7,k,n,p,q,g,h}$ is stored and processed on the host $R_{13}$. Other forms of mapping can be accomplished,
since $G_{ij}$ is used  as a load distribution  unit.

\begin{figure}[!hb]
 \centering
 \includegraphics[bb=0 0 478 371,scale=0.55]{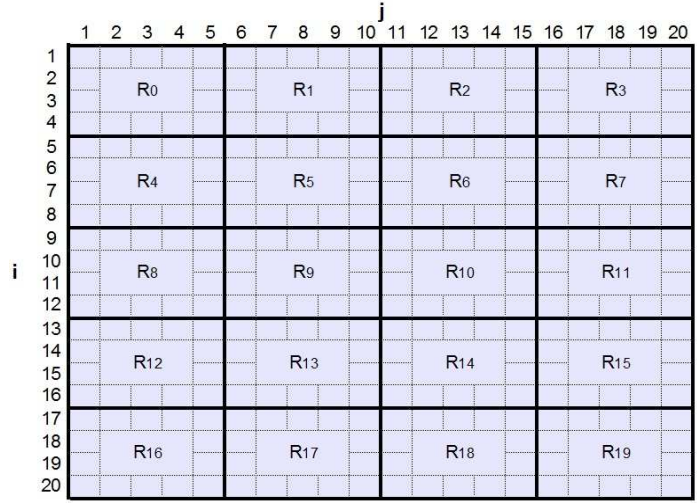}
 \label{figura1}
 \caption {Example of allocation of  sets  $G_{ij}$ on  20 hosts}
\end{figure}

The RLT3 algorithm applied to the QAP requires a lot of RAM memory  to store the coefficients of the matrices. 
An instance with  $N=30$, for example, requires  around 1.6 TByte to store the matrix $E$, which is  composed of $N^2$ x $(N -1)^2$ x $(N -2)^2$ x $(N -3)^2$
 elements, each one  keeping  an integer or float  data (4 bytes). 
Although some improvements  have been proposed  in \citet{hahn2012}, the required memory  goes on being  much bigger than the provided  by  modern computers.

In the distributed algorithm, complementaries belonging to different sets can be allocated on different hosts, requiring that hosts communicate 
among themselves during their executions.  
 The distributed algorithm runs several iterations and at each of them, four steps are executed. 
In the first one, complementary costs of matrix $E$ are exchanged. Complementary costs stored in $R_x$, needed in $R_z$, are transferred through messages from $R_x$ to $R_z$, denoted as $Comp(E)_{xz}$. 
In the next two steps, complementary costs of matrices $D$ and $C$,  are sent through  $Comp(D)_{xz}\ and\ Comp(C)_{xz}$ messages, respectively.
In the  final stage, matrices $B$ are transmitted through $Mens(B)_{xz}$ messages. 
In small instances,  up to $N=12$, communication overhead does not impact the performance negatively. However, in bigger instances,  the communication of complementary costs of matrix E can represent 
 up to  70\% of the total execution time in instances with $N = 30$.\

\

The steps of the distributed algorithm executed in the process $R_t$ are described next.

\noindent
{\bf 1- Initialization:}  $LB \leftarrow 0$,   $B_{ij} \leftarrow 0\ \forall \ (i,j)$,      
 $C_{ijkn} \leftarrow$ $f_{ik} \times d_{jn}$ $\ \forall \ (i,j,k,n)$ with $i\not=k\ $ and $\ j\not=n$,
 $D_{ijknpq} \leftarrow 0$ $\ \forall \ (i,j,k,n,p,q)$ with $ i\not=k\not=p\ $ and $ \ j\not=n\not=q$,
 $E_{ijknpqgh} \leftarrow 0 $ $\ \forall \ (i,j,k,n,p,q,g,h)$ with $i\not=k\not=p\not=g$ and $j\not=n\not=q\not=h$,  
$\ cont \leftarrow 1$, 
$\ lim \leftarrow $ total of iterations and  $optimal \leftarrow$ optimal solution or best known solution cost.\

\noindent
{\bf 2 - Transferring   complementaries of matrix $C$:} For each $ R_s \in T\ $ and $\ R_s \not= R_t$, and for each $(i,j,k,n)\ |\ 
G_{ij}$ allocated in $R_t\ and \ G_{kn}$ allocated in $R_s,\ i < k \ $ and $\ j\not=n$, 
store coefficients $C_{ijkn}$ in $Comp(C)_{ts}$ $\ \forall \  i < k \ $ and $\ j\not=n$. Send $Comp(C)_{ts}$  to   $R_s$. 
Upon receiving messages from other hosts,  for each $G_{ij}$ allocated in $R_t $,\ $C_{ijkn} \leftarrow (C_{ijkn} + C_{knij} )/2$. \ \

\noindent
{\bf 3 -  Cost concentration from matrix $C$ to matrix  $B$:} For each $G_{ij}$ allocated in  $R_t $, concentrate  the  coefficients from 
matrix \ $C$ \ to \ $B$, \ by executing the \ Hungarian \ Algorithm, \  $B_{ij}$ $\leftarrow  Hungarian(C_{ij})$.\

\noindent
{\bf 4- Transferring  matrix $B$:} For each $(i,j)\ |\  G_{ij}$ allocated in  $R_t $, 
store coefficients $B_{ij}$ in $Mens(B)$.   Broadcast $Mens(B)$ to all hosts.
After receiving messages from all other hosts, update local matrix $B$.

\noindent
{\bf 5- Cost concentration from matrix $B$ to $LB$:} $ LB \leftarrow  Hungarian(B)$.

\noindent
{\bf 6 - Loop:} Repeat until $cont = lim$ or $LB = optimal$, 
 The loop termination condition is  achieved when the total number of iterations reaches the previously defined
 limit ($cont = lim$) or the optimal solution is equal to the current lower bound ($LB = optimal$).

\noindent
{\bf 7 - Cost spreading from matrix $B$ to $C$:}  
For each $(i,j)\ |\ G_{ij}$ allocated in $R_t $, 
spread $B_{ij}$ through $(N-1)$ submatrix rows of  $C_{ij}$. 
Each cost element $C_{ijkn}$ is increased by $B_{ij}\  /\ (N-1)$ $\forall \ k \not=i\ e\ j \not=n$.

\noindent
{\bf 8 - Cost spreading from matrix $C$ to $D$:}  
For each $(i,j,k,n) \ |\ G_{ij}$ allocated in $R_t $ and $i \not= k\ $ and $\ j \not= n $, 
spread $C_{ijkn}$ through $(N-2)$ submatrix rows of $D_{ijkn}$.
Each cost element $D_{ijknpq}$ is increased by $C_{ijkn}\ /\ (N-2)$ $\forall \ p \not=i,k\ $ and $\ q \not=j,n$. 

\noindent
{\bf 9 - Cost spreading from matrix $D$ to $E$:} 
For each $(i,j,k,n,p,q) \ |\ G_{ij}$ allocated in $R_t $ and $i \not= k,p\ $ and $\ j \not= n,q$,  
spread $D_{ijknpq}$ through $(N-3)$ submatrix rows of $E_{ijknpq}$.
Each cost element $E_{ijknpqgh}$ is increased by $D_{ijknpq}\ /\ (N-3)\ \forall \ g \not=i,k,p\ $ and $\ h \not=j,n,q$. 

\noindent
{\bf 10 - Cost transfer between complementary coefficients of matrix $E$:}
For each $R_s \in T \ $ and $\ R_s \not= R_t$, for each $(i,j,k,n,p,q,g,h)\ |\
G_{ij}$ allocated in $ R_t\ $ and $\ (G_{kn},G_{pq}$ or $G_{gh})$ allocated in $R_s$ and $ i < k < p < g\ $
and $\ j\not=n\not=q\not=h$, include the coefficients $E_{ijknpqgh}$ in $Comp(E)_{ts}$.
Send message containing $Comp(E)_{ts}$.
Upon receiving messages from all hosts, for each $(i,j,k,n,p,q,g,h)\ |\ G_{ij}$ allocated in $R_t $,\ 
$E_{ijknpqgh} \leftarrow E_{ijknghpq} \leftarrow E_{ijpqkngh} \leftarrow E_{ijpqghkn}  \leftarrow E_{ijghknpq} \leftarrow E_{ijghpqkn}  
\leftarrow  (E_{ijknpqgh} + E_{ijknghpq} +\ ...\ + E_{ghpqknij})/24$. \ 

\noindent
{\bf 11 - Cost concentration from matrix $E$ to $D$:} For each $(i,j,k,n,p,q)\ |\ G_{ij}$ allocated in $R_t $, concentrate 
the submatrices from $E$ to $D$, i.e., $D_{ijknpq}$ $\leftarrow  Hungarian(E_{ijknpq})$.

\noindent
{\bf 12 - Cost transfer between complementary coefficients of matrix $D$:}
For each $R_s \in T\ $ and $\ R_s \not= R_t$, for each $(i,j,k,n,p,q)\ |\ 
G_{ij}$ allocated in $R_r\ $ and $\ (G_{kn}$ or $G_{pq})$ allocated in $R_s\ $ and $\ i < k < p\ $ and $ \ j\not=n\not=q,$
include the coefficients $D_{ijknpq}$ in $Comp(D)_{ts}$. Send message containing $Comp(D)_{ts}$.
Upon receiving messages from all hosts, for each $(i,j,k,n,p,q)\ |\ G_{ij} \in R_t $,\ $D_{ijknpq} \leftarrow D_{ijpqkn} \leftarrow
(D_{ijknpq} + D_{ijpqkn} + D_{knijpq} + D_{knpqij} + D_{pqijkn} + D_{pqknij})/6$.

\noindent
{\bf 13 - Cost concentration from matrix $D$ to $C$:} For each $(i,j,k,n)\ |\ G_{ij}$ allocated in $R_t $, concentrate 
the submatrices from $D$ to $C$, i.e. , $C_{ijkn}$ $\leftarrow  Hungarian(D_{ijkn})$.

\noindent
{\bf 14, 15, 16, and 17 - These steps are identical to Steps 2, 3, 4, and 5, respectively.}

\noindent
{\bf 18 - loop end:} Increase the variable $cont$ and return to Step 6.

Compared to the sequential version, the following modifications have been applied in the distributed algorithm:
(i) use of floating point numbers instead of integers for cost coefficients;
(ii) use of arithmetic means to transfer costs among complementary coefficients;
(iii) execution of all cost transfers among complementary coefficients before concentration; and
(iv) never spreading from $LB$ to matrix $B$.

From all these differences, the most important one is that of item (ii).
In the sequential dual ascent algorithm proposed in \citet{hahn2012}, cost transfers are performed with the
aim of increasing all cost coefficients of the current submatrix $M$, by pushing residual cost from its
complementaries, before applying the cost concentration in that matrix.
This approach imposes a sequential handling of submatrices at the same RLT level.
Taking arithmetic means allow that such matrices are processed in parallel but prevents from using residual
costs resulting from the Hungarian algorithm in other matrices at the same RLT level in the same iteration.
This {\it reuse} of costs is not possible because all costs are evenly distributed among all complementaries
before all cost concentrations are performed at that level.
Initially, we expected that such modification would significantly slow down the convergence of the lower bound
and/or substantially reduce its quality but the experiments reported in the next section show that neither
effects are observed.
In fact, we obtained better lower bounds in some cases.

\section{Experimental Results} \label{sec:resultados}

\begin{table}[!htb]
\scriptsize
\caption{ Comparison between the newly proposed distributed algorithm and other techniques }
\centering
\begin{tabular}{l|c|c|c|c|c|c|c|c|c|c} \hline
         &         &              &               &                 & \multicolumn{6}{c}{Distributed}  \\            
Instance & Optimal & BV04         & HH01          & HZ07            & \multicolumn{6}{c}{Version}  \\ \cline{6-11}
         &         &              &               &                 & $LB$          &  \textit{gap}   & time(s)  & Speedup & hosts & iterations   \\    \hline \hline
had14    & 2724    & {\bf 0.00\%*} &  -            & -              &  2724       & {\bf 0.00\% *} & 559       &  1.62   &  4    & 29 \\ 
had16    & 3720    & 0.13\%       & {\bf 0.00\% *} & 0.02\%         &  3720       & {\bf 0.00\% *} & 744       &  5.83   &  8    & 22  \\ 
had18    & 5358    & 0.11\%       & {\bf 0.00\% *} & 0.02\%         &  5358       & {\bf 0.00\% *} & 5456      &  5.27   &  9    & 59 \\ 
had20    & 6922    & 0.16\%       & {\bf 0.00\% *} & 0.03\%         &  6922       & {\bf 0.00\% *} & 16118     &   NA    & 16    & 109  \\ \hline 
kra30a   & 88900   & 2.50\%       & 2,98\%         & -              &  88424      & {\bf 0.54\% }  & 196835    &   NA    & 90    & 162 \\ \hline
nug12    & 578     & 1.73\%       & {\bf 0.00\% *} & 0.14\%         &  578        & {\bf 0.00\% *} & 73        &  2.75   &  4    & 16 \\ 
nug15    & 1150    & 0.78\%       & {\bf 0.00\% *} & 0.08\%         &  1150       & {\bf 0.00\% *} & 360       &  5.28   &  9    & 22  \\  
nug16a   & 1610    & 0.75\%       & -             &  -              &  1610       & {\bf 0.00\% *} & 1132      &  5.73   &  8    & 34 \\ 
nug16b   & 1240    & 1.69\%       & -             &  -              &  1240       & {\bf 0.00\% *} & 1294      &  5.71   &  8    & 39 \\ 
nug18    & 1930    & 1.92\%       & -             & {\bf 0.00\% *}  &  1930       & {\bf 0.00\% *} & 7172      &  5.36   &  9    & 78 \\ 
nug20    & 2570    & 2.49\%       & 2.41\%        & 0.14\%          &  2570       & {\bf 0.00\% *} & 30129     &   NA    & 20    & 249 \\ 
nug22    & 3596    & 2.34\%       & 2.36\%        & 0.08\%          &  3596       & {\bf 0.00\% *} & 41616     &   NA    & 22    & 157 \\ 
nug24    & 3488    & 2.61\%       & -             &  -              &  3478       & {\bf 0.28\% }  & 173520    &   NA    & 24    & 300 \\
nug25    & 3744    & 3.29\%       & -             &  -              &  3689       & {\bf 1.44\% }  & 172020    &   NA    & 25    & 211 \\ 
nug28    & 5166    & 2.92\%       & -             &  -              &  5038       & {\bf 2.48\% }  & 171783    &   NA    & 49    & 118\\ 
nug30    & 6124    & 3.10\%       & 5.78\%        &  -              &  5940       & {\bf 3.00\% }  & 229583    &   NA    & 100   & 119 \\ \hline
rou15    & 354210  & 1.13\%       & {\bf 0.00\% *}& {\bf 0.00\% *}  &  354210     & {\bf 0.00\% *} & 323       &  5.78   & 9     & 20  \\ 
rou20    & 725520  & 4.19\%       & 3.60\%        & {\bf 0.03\%  }  &  720137     & 0.74\%         & 37079     &   NA    & 25    & 300 \\ \hline
tai15a   & 388214  & 2.86\%       & -             &  -              &  388214     & {\bf 0.00\% *} & 737       &  6.18   &  9    &  46 \\ 
tai17a   & 491812  & 3.11\%       & -             &  -              &  491812     & {\bf 0.00\% *} & 1259      &  13.18  & 17    &  46 \\ 
tai20a   & 703482  & 4.52\%       & 3.93\%        & {\bf 703482 * } &  698271     & 0.74\%         & 45720     &   NA    & 25    & 300 \\ 
tai25a   & 1167256 & 4.66\%       & 6.48\%        &  -              &  1122200    & {\bf 3.87\% }  & 101170    &   NA    &  25   & 124   \\ 
tai30a   & 1818146 & 6.12\%       & 7.25\%        &  -              &  1724510    & {\bf 5.15\% }  & 112085    &   NA    & 100   & 58 \\ \hline
tho30    & 149936  & 4.75\%       & 9.82\%        &  -              &  142990     & {\bf 4.63\% }  & 145713    &   NA    & 100   & 79 \\ \hline
chr18a   & 11098   & {\bf 0.00\% *}& -            &  -              &  11098      & {\bf 0.00\% *} & 1892      &  5.32   &   9   & 20  \\ 
chr20a   & 2192    & 0.18\%       & -             &  -              &  2192       & {\bf 0.00\% *} & 5914      &   NA    &  16   & 39   \\ 
chr20b   & 2298    & 0.13\%       & -             &  -              &  2298       & {\bf 0.00\% *} & 3708      &   NA    &  16   & 24   \\ 
chr22a   & 6156    & 0.03\%       & -             &  -              &  6156       & {\bf 0.00\% *} & 5321      &   NA    &  22   & 20  \\ 

\end{tabular}
\label{Tabela1}
\end{table}

The application was implemented using the programming language C++ with the library IntelMPI library.
The experiments were performed in the 
Netuno Cluster, see \citet{netuno2011}, a cluster composed of 256 hosts, interconnected by infiniband.
Each host consists of a two Intel Xeon E5430 2.66GHz
Quad core processor with 12MB cache L2 and 16 GB of RAM per host. 

A unique process is executed per host, allowing that it  uses  the total available memory  without resource contention
usually caused by process concurrency. So,  only one core per host is used to execute the application.   
 
For evaluation of the proposed distributed algorithm, the application terminates when the optimal solution
is found or when a total of 300 iterations is executed, respecting a time limit (usually about three days
per instance) that varies according the machine availability in the cluster.

Table \ref{Tabela1} presents the results for different instances and sizes from the QAPLIB.
In the first column of Table \ref{Tabela1}, there are the instance names and the corresponding dimensions.
For example, \textit{nug20} represents an instance \textit{nug}, from  \citet{nug68}, with size $N=20$. 
In the second column, there are the optimal values for each instance.
The third column (BV04) contains the gaps obtained by the lift-and-project relaxation proposed in
\citet{burer2006}.
At the fourth column (HH01), one finds the gaps obtained by the RLT2 based dual ascent algorithm proposed
in \citet{adams2007}. 
In the fifth column (HZ07), there are the gaps obtained by the RLT3 based dual ascent algorithm proposed
in \citet{hahn2012}.
The results presented for the last two methods were obtained from the QAPLIB website, which does not contain
values for all instances.
In the sixth column, we show the lower bounds obtained in the RLT3 distributed version proposed in this paper.
In the seventh column, we present the corresponding gaps, in the eighth column, the execution times in seconds,
and in the last three columns, the speedups obtained via parallelism, the number of hosts used, and the number
of iterations performed.

Also in Table \ref{Tabela1}, notice that the lower bounds that correspond to optimal solution costs or gaps
that are zero are marked with an asterisk, and those which are the best known gaps are in bold printed.
For some instances, it was not possible to execute the sequential versions because of the memory constraints, in those cases the calculation of speedups were not applicable, as indicated in the table (NA).

\section{Conclusion} \label{sec:conclusao}

The distributed version achieved goods results compared with other proposals, reaching the best known bounds of 26 out of 28 instances, 
being  18 of them the optimal solutions.
The distributed algorithm allowed the execution of instances with size $N=28$ and $N=30$ for the first time using RLT3. Those good results were achieved due to the use some of parallelism and the changes proposed in the original sequential code.

\bibliographystyle{model5-names}
\bibliography{ref}

\end{document}